# Quantum Cellular Automata

by

Bassam Aoun

&

Mohamad Tarifi

# ABSTRACT

We provide an introduction to Quantum Cellular Automata.

# ACKNOWLEDGEMENTS

*We acknowledge our friend Carlos for extremely useful conversations and his good sense of humor!*

*We acknowledge Prof. Michel Mosca for his excellent course that introduced and hooked us to Quantum Information Processing.*

*We also acknowledge University of Waterloo for being a good host.*

# CONTENTS

# 1. INTRODUCTION

*I do not know where to start, so I will just state the obvious hopping that this will let us realize some of the obviousness that we miss.*

The emergence of simple patterns out of complex systems motivates the study of behavior independent of the particulars of a system. Cellular Automata is a simple tool that displays such characteristics and is therefore useful for modeling.

Cellular automata are particularly useful for presenting parallel computation, and can be thought of as its basic building blocks. Cellular automata sometimes display complex behavior even when simple rules are applied.

It is natural then, to extend the models of cellular automata to encompass what we believe about nature and computation.

In this report we attempt to provide a useful introduction to quantum cellular automata from a computing perspective. For clarity and accessibility we provide a brief overview of both quantum computing and classical cellular automata.

## 2. BRIEF OVERVIEW OF QUANTUM MECHANICS

### 2.1 THE BEAM SPLITTER EXPERIMENT

As a start, we will present two brief experiments to illustrate some basic concepts in quantum mechanics. The outcome of the last experiment seems counterintuitive because everyday phenomena are well within good classical physics approximation range.

In figure 1, a light source emits a photon along a path towards a half-silvered mirror. This mirror splits the light, reflecting half vertically towards detector A and transmitting half toward detector B. Our intuition would tell us that the photon leaves the mirror either towards A or B with equal probability since it cannot be split. The fact that a photon cannot split have been verified through detecting a signal at only one detector. This means that photons will be detected 50% of the time at each of the two detectors. So far, the quantum physical prediction agrees with the classical one.

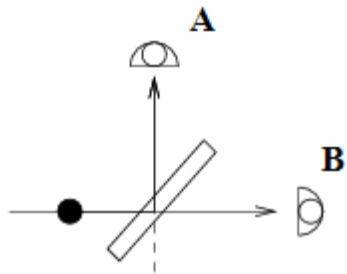

**Figure 1: Experiment 1 using one beam splitter**

This peace of information is misleading since it might lead us to think that the photon leaves either towards A or towards B. However, quantum mechanics predicts, through the effect known as *single-particle interference*, that the photon actually travels along both paths simultaneously, collapsing down to one path only upon measurement. The following experiment illustrates the last effect.

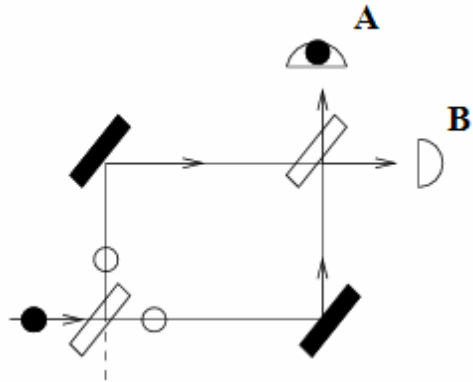

**Figure 2: Experiment 2 using two beam splitters**

In this experiment, we introduce a fully silvered mirror instead of each detector of the first experiment such that the two paths will encounter a half-silvered mirror before reaching detectors A and B. Once a photon will reach the last half-silvered mirror, along either one of the two paths, one might think that the photon will reach the detectors A or B with a probability of 50% each. However in this experiment, the detector A or the detector B will register a hit 100% of the time whereas the other one will never be triggered.

In this experiment, our classical intuition based on the conditional probability doesn't predict such outcome. We cannot explain this conclusion based on a comparison with the first experiment. This phenomenon is known as single-particle *interference.* One interpretation quantum physics states that the photon traveled both paths simultaneously; creating interference at the point of intersection that canceled the possibility of the photon to reach the other detector. Consequently, if we cancel out the effect of quantum interference by placing an absorbing screen on one of the paths, both detectors will registers 50% hits similar to the first experiment. Those potential paths taken by the photon represent the superposition of the possible photon states.

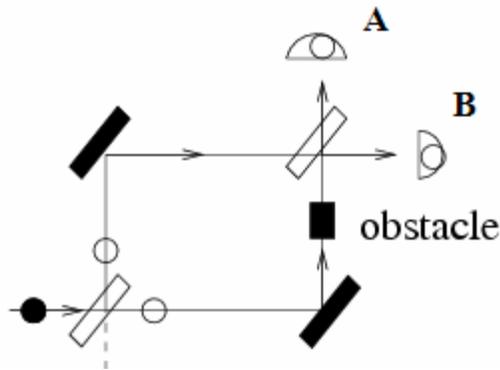

**Figure 3: Placing an obstacle on one of the paths**

Those special characteristics as the superposition of different states and interference give the quantum computer the potential to be incredible powerful computational devices. Therefore, quantum computers are not seen as continuity of classical computers but as an entirely new branch of thoughts.

2.2 QUANTUM MECHANICS POSTULATES

In this section we introduce the four postulates of quantum mechanics as they are relevant to our investigation in quantum information processing. Quantum postulates are very important in a sense that they provide the connections between the physical, real, world and the quantum mechanics mathematics used to model these systems.

**Postulate 1**: *Any isolated physical space is associated with a complex vector space with inner product called the State Space of the system.* It states that a system is completely described by a state vector, a unit vector, pertaining to the state space which describes all the possible states the system can be in.

**Postulate 2:** *Evolution of an isolated system can be as:*

$$|v(t_2)\rangle = U(t_1, t_2)|v(t_1)\rangle$$

*where t1, t2 are moments in time and U(t1, t2) is a unitary operator*. We should note that the process is reversible, since

$$U^\dagger U |v\rangle = |v\rangle$$

**Postulate 3**: *The measurement of a closed system is described by a collection of operators Mm which act on the state space such that*

- $\rho(m) = \langle \psi | M_m^\dagger M_m | \psi \rangle$ *describes the probability the measurement outcome* **m** *occurred,*

- $|\psi'\rangle = \dfrac{M_m |\psi\rangle}{\sqrt{\langle \psi | M_m^\dagger M_m | \psi \rangle}}$ *is the state of the system after measurement outcome* **m** *occurred,*

- $\sum_m M_m^\dagger M_m = I \Leftrightarrow \sum_m \rho(m) = 1$ *(completeness relation).*

Note that measurement is an external observation of a system and so disturbs its unitary evolution.

**Postulate 4:** *The state space of a composite system is the tensor product of the state spaces of its components*

$$\left.\begin{array}{l} System\ A:\ |x\rangle \\ System\ B:\ |\psi\rangle \end{array}\right\} System\ AB:\ |x\rangle \otimes |\psi\rangle.$$

However, if the qubits are allowed to interact, then the closed system includes all the qubits together, and it may not be possible to write the state in the product form. When this is the case, we say that the qubits in the ensemble are *entangled* (refer to later sections for further analysis of entanglement).

2.3 QUANTUM BITS

*2.3.1Qubits*

The fundamental resource and basic unit of quantum information is the quantum bit (qubit). From a physical point of view, a qubit is represented by an ideal two-state quantum system. Examples of such systems include photons (vertical and horizontal polarization), electrons and other spin-1/2 systems (spin up and down), and systems defined by two energy levels of atoms or ions. From the beginning the two-state system played a central role in studies of quantum mechanics. It is the most simple quantum system, and in principle all other quantum systems can be modeled in the state space of collections of qubits.

A qubit is represented as unit vector in a two dimensional complex vector space for which a particular orthonormal basis, denoted by $\{|0\rangle, |1\rangle\}$, has been fixed. The notation for these states was introduced by Dirac. It is called the "ket" notation, and its variations are widely used in quantum physics. It is important to notice that the basis vector $|0\rangle$ is not the zero vector of the vector space.

For the purposes of quantum computing, the basis states $|0\rangle$ and $|1\rangle$ encode the classical bit values 0 and 1 respectively. Unlike classical bits however, qubits can be in a superposition of $|0\rangle$ and $|1\rangle$ such as $\alpha|0\rangle + \beta|1\rangle$ where α and β are complex numbers such that $|\alpha|^2 + |\beta|^2 = 1$. If such a superposition is measured with respect to the basis $\{|0\rangle, |1\rangle\}$, the probability that the measured value is $|0\rangle$ is $|\alpha|^2$ and the probability that the measured value is $|1\rangle$ is $|\beta|^2$.

Key properties of quantum bits:

1. A qubit can be in a superposition state of 0 and 1.

2. Measurement of a qubit in a superposition state will yield probabilistic results.

3. Measurement of a qubit changes the state to the result.

*2.3.2 Tensor products*

Much computational power of quantum systems comes from the fact that as the number of qubits increases linearly, the amount of information stored increases exponentially. For example, a single-qubit state $|\psi\rangle \in \mathbb{C}^2$ is represented by a pair of complex numbers: $|\psi\rangle = \alpha|0\rangle + \beta|1\rangle$. The composite state of two qubits is an element of $\mathbb{C}^4$:

$$\alpha_{00}|00\rangle + \alpha_{01}|01\rangle + \alpha_{10}|10\rangle + \alpha_{11}|11\rangle.$$

The composite state of three qubits is in $\mathbb{C}^8$, and so on.

More generally, if $H_1$ and $H_2$ are Hilbert spaces, then $H_1 \otimes H_2$ is also a Hilbert space. If $H_1$ and $H_2$ are finite dimensional with bases $\{u_1, u_2, ... u_n\}$ and $\{v_1, v_2, ... v_n\}$ respectively, then $H_1 \otimes H_2$ has dimension *nm* with basis $\{u_i \otimes v_j \mid 1 \leq i \leq n, 1 \leq j \leq m\}$.

For matrices $A$, $B$, $C$, $D$, $U$ and scalars $a$, $b$, $c$, $d$ the following hold:

$$\begin{pmatrix} A & B \\ C & D \end{pmatrix} \otimes U = \begin{pmatrix} A \otimes U & B \otimes U \\ C \otimes U & D \otimes U \end{pmatrix} \quad \text{and} \quad \begin{pmatrix} a & b \\ c & d \end{pmatrix} \otimes U = \begin{pmatrix} a \otimes U & b \otimes U \\ c \otimes U & d \otimes U \end{pmatrix}$$

The tensor product of several matrices is unitary if and only if each one of the matrices is unitary up to a constant. Let $U = A_1 \otimes ... \otimes A_n$. Then $U$ is unitary if $A_i^\dagger A_i = k_i I$ and $\prod_i k_i = 1$.

$$U^\dagger U = (A_1^\dagger \otimes ... \otimes A_n^\dagger)(A_1 \otimes ... \otimes A_n) = A_1^\dagger A_1 \otimes ... \otimes A_n^\dagger A_n = k_1 I \otimes ... \otimes k_n I = I$$

We can define an inner product on $U \otimes V$ by

$$(|v_1\rangle \otimes |u_1\rangle, |v_2\rangle \otimes |u_2\rangle) = (|v_1\rangle, |v_2\rangle).(|u_1\rangle, |u_2\rangle),$$

which could be written in another notation as

$$\langle v_1 \otimes u_1 | v_2 \otimes u_2 \rangle = \langle v_1 | v_2 \rangle \langle u_1 | u_2 \rangle,$$

*2.3.3 Entangled quantum states*

The fundamental observation of Josza R., in "Entanglement and quantum computation", states that entanglement, not superposition, is the essential feature that empowers quantum computation, and is what gives other quantum technologies (such as quantum teleportation) their power.

A classical (macroscopic) physical object broken into pieces can be described and measured as separate components. An *n*-particle quantum system cannot always be described in terms of the states of its component pieces. For instance, the state $|00\rangle + |11\rangle$ cannot be decomposed into separate states of each of the two qubits in the form $(a_1|0\rangle, b_1|1\rangle) \otimes (a_2|0\rangle + b_2|1\rangle)$. This is because $(a_1|0\rangle, b_1|1\rangle) \otimes (a_2|0\rangle + b_2|1\rangle) = a_1 a_2 |00\rangle + a_1 b_2 |01\rangle + b_1 a_2 |10\rangle, b_1 b_2 |11\rangle$ and $a_1 b_2 = 0$ implies that either $a_1 a_2 = 0$ or $b_1 b_2 = 0$. States which cannot be decomposed in this way are called entangled states. These are states that don't have classical counterparts, and for which our intuition is likely to fail.

When particles are entangled, a measurement of one affects a measurement of the other. For example, the state $\frac{1}{\sqrt{2}}(|00\rangle + |11\rangle)$ is entangled since the probability of measuring the first bit as $|0\rangle$ is $1/2$ if the second bit has not been measured. However, if the second bit has been measured, the probability that the first bit is measured as $|0\rangle$ is either 1 or 0, depending on whether the second bit was measured as $|0\rangle$ or $|1\rangle$, respectively. On the other hand, the state $\frac{1}{\sqrt{2}}(|00\rangle + |01\rangle)$ is not entangled. Since $\frac{1}{\sqrt{2}}(|00\rangle + |01\rangle) = |0\rangle \otimes \frac{1}{\sqrt{2}}(|0\rangle + |1\rangle)$, any measurement of the first bit will yield $|0\rangle$ regardless of measurements of the second bit. Similarly, the second bit has a fifty-fifty chance of being measured as $|0\rangle$ regardless of measurements of the first bit. Therefore, entanglement is a non-classical correlation between two quantum systems. It is most strongly exhibited by the maximally entangled states such as the Bell states for two qubits, and is considered to be absent in mixtures of product states ("separable" states). Often states that are not separable are considered to be entangled. However, nearly separable states do not exhibit all the features of maximally entangled states. As a result, studies of different types of entanglement are an important component of quantum information theory.

## 2.4 QUANTUM COMPUTING

This exponential growth in number of states, together with the ability to subject the entire space to transformations (either unitary dynamical evolution of the system, or a measurement projection into an eigenvector subspace), provides the foundation for quantum computing.

An interesting (apparent) dilemma is the energetic costs/ irreversibility of classical computing. Since unitary transformations are invertible, quantum computations (except measurements) will all be reversible by restricting them to unitary quantum transformations. This means that every quantum gate (on one or many qubits) implements a reversible computation. That is, given the output of the gate, it must be possible to unambiguously determine what the input was. Fortunately, there is a classical theory of reversible computation that tells us that every classical algorithm can be made reversible with an acceptable overhead, so this restriction on quantum computation does not pose a serious problem. It is something that must be kept in mind when proposing a specification for a quantum gate, however.

We will illustrate quantum gate representation through an example of the quantum version of the classical *not* gate. It is represented by $\sigma_x$ and has the effect of negating the values of the computational basis. That is, using *ket* notation,

$$not(\alpha|0\rangle + \beta|1\rangle) = \alpha|1\rangle + \beta|0\rangle = \beta|0\rangle + \alpha|1\rangle$$

In vector notation this equation becomes: $not\begin{pmatrix}\alpha\\\beta\end{pmatrix} = \begin{pmatrix}\beta\\\alpha\end{pmatrix}$.

Another effect of expressing the effect of *not* is by multiplying the vector by a matrix representing *not*:

$$not\begin{pmatrix}\alpha\\\beta\end{pmatrix} = \begin{pmatrix}0 & 1\\1 & 0\end{pmatrix}\begin{pmatrix}\alpha\\\beta\end{pmatrix} = \begin{pmatrix}\beta\\\alpha\end{pmatrix}$$

so we can identify the action of not with the matrix $\sigma_x = \begin{pmatrix}0 & 1\\1 & 0\end{pmatrix}$.

Similarly, we can find some useful single-qubit quantum state transformations. Because of linearity, the transformations are fully specified by their effect on the basis vectors. The associated matrix is also shown. They are known as the four the four Pauli gates.

$$I : \begin{array}{c} |0\rangle \to |0\rangle \\ |1\rangle \to |1\rangle \end{array} \begin{pmatrix} 1 & 0 \\ 0 & 1 \end{pmatrix}$$

$$\sigma_y : \begin{array}{c} |0\rangle \to |1\rangle \\ |1\rangle \to -|0\rangle \end{array} \begin{pmatrix} 0 & -1 \\ 1 & 0 \end{pmatrix}$$

$$\sigma_z : \begin{array}{c} |0\rangle \to |0\rangle \\ |1\rangle \to -|1\rangle \end{array} \begin{pmatrix} 1 & 0 \\ 0 & -1 \end{pmatrix}$$

Note that $I$ is the identity transformation (often called *nop* or no-operation), $\sigma_x$ is negation, $\sigma_z$ is a phase shift operation, and $\sigma_y = \sigma_z \sigma_x$ is a combination of both. One reason why the Pauli gates are important for quantum computing is that they span the vector space formed by all 1-qubit operators.

All these gates are indeed unitary. For example:

$$\sigma_y \sigma_y^\dagger = \begin{pmatrix} 0 & -1 \\ 1 & 0 \end{pmatrix} \begin{pmatrix} 0 & 1 \\ -1 & 0 \end{pmatrix}$$

Another very important gate is the Hadamard gate defined by the following transformation:

$$H : \begin{array}{c} |0\rangle \to |0\rangle + |1\rangle \\ |1\rangle \to |0\rangle - |1\rangle \end{array}$$

Applied to $n$ bits each in the $|0\rangle$ state, the transformation generates a superposition of all $2^n$ possible states.

$$H^{\otimes n} |0\rangle^{\otimes n} = \frac{1}{\sqrt{2^n}} ((|0\rangle + |1\rangle) \otimes \ldots \text{n times} \ldots \otimes (|0\rangle + |1\rangle)) = \frac{1}{\sqrt{2^n}} \sum_{x=0}^{2^n - 1} |x\rangle$$

Other then the Hadamard gate, we need to mention the T gate. It is sometimes referred to as the $\frac{\pi}{8}$ gate. It is represented by the following matrix:

$$T = \begin{pmatrix} 1 & 0 \\ 0 & e^{i\frac{\pi}{4}} \end{pmatrix} = e^{i\frac{\pi}{8}} \begin{pmatrix} e^{-i\frac{\pi}{8}} & 0 \\ 0 & e^{i\frac{\pi}{8}} \end{pmatrix}$$

An important two-qubit operator is the *CNOT*. It is given as follows:

$$CNOT|00\rangle = |00\rangle$$
$$CNOT|01\rangle = |01\rangle$$
$$CNOT|10\rangle = |11\rangle$$
$$CNOT|11\rangle = |10\rangle$$

Classically, we can think of the *C-not* as flipping the second register if and only if the first register is set to 1. The transformation $C_{not}$ is unitary since $C_{not}^{\dagger} = C_{not}$ and $C_{not} C_{not} = I$. The $C_{not}$ gate cannot be decomposed into a tensor product of two single-bit transformations.

By analogy to classical computation, one may ask what kind of quantum gates we need in order implement an arbitrary unitary transformation. Since the number of possible unitary transformation is uncountable, one can not find a set of basic gates that construct exactly every unitary transformation. However, there exist universal gates which can construct any transformation U within bounded error ε. Said in other words, we can construct a circuit U' from those gates that simulates U within the allowed ε. A universal set of operations is: H, X, T, and $C_{Not}$.

## 2.5 QUNAUTM ALGORITHM

Quantum algorithms are methods using quantum networks and processors to solve algorithmic problems. On a more technical level, a design of a quantum algorithm can be seen as a process of an efficient decomposition of a complex unitary transformation into products of elementary unitary operations (or gates), performing simple local changes.

The four main properties of quantum mechanics that are exploited in quantum computations are:

- Superposition
- Interference
- Entanglement
- Measurement

The publication of P. Shor's quantum algorithm in 1994 for efficiently factoring numbers was a key event that stimulated many theoretical and experimental investigations of quantum computation. One of the reasons why this algorithm is so important is that the security of widely used public key cryptographic protocols relies on the conjectured difficulty of factoring large numbers. Furthermore, more recently, Lov Grover came up with a quantum search algorithm that has a run time complexity proportional to $\sqrt{n}$ without any knowledge about the function f. Even though it is not an exponential speed-up, it is an improvement over classical search algorithms.

## 2.6 QUANTUM TURING MACHINES

We discuss the 'automata theoretic' definition due to Bernstein, Vazirani and Deutsch, an alternative and more 'physical' approach is given by Benioff.

Quantum Turing Machines (QTMs) are analogous to probabilistic Turing Machines. As always, QTMs are made of a processing unit, a tape divided into discrete cells, a read/write head, and a set of states.

More formally, a single tape quantum Turing machine is a quintuple $\langle \Sigma, Q, q_0, q_f, \delta \rangle$, where $\Sigma$ is the finite tape alphabet that includes the blank symbol, $Q$ is a finite set of states, $q_0$ and $q_f$ are the initial and final states respectively.

The transition function $\delta$ maps:

$$\delta : (Q \times \Sigma \times \Sigma \times Q \times \{\leftarrow, \downarrow, \rightarrow\}) \to \mathbb{C}_{[0,1]}$$

As usual the number of non-blank symbols is assumed to be finite.

A configuration of a QTM is a triplet $\langle \tau, i, q \rangle$, where $\tau$ is content of the tape, $i$ specifies the position of the head and $q$ the current state of machine.

If we let C denote the set of all such configurations, then computation is done in the inner product space $H = l_2(C)$ with bases $\{|c\rangle \mid c \in C\}$. We can derive a mapping $a : C \times C \to \mathbb{C}$ from the transition function such that for $c_1, c_2 \subset C$, $a(c_1, c_2)$ is the amplitude of the transition from the basis state $|c_1\rangle$ to $|c_2\rangle$.

The time evolution behavior of the QTM can then be defined as a mapping $U : H \to H$ that takes a superposition of configurations $|\phi\rangle = \sum_{c \in C} \alpha_c |c\rangle$ maps to $U|\phi\rangle = \sum_{c \in C} \alpha_c U|c\rangle$, where $U|c\rangle = \sum_{c' \in C} a(c, c') |c'\rangle$.

If one measures the state $|\phi\rangle$ of QTM it will collapse to a TM state $|c\rangle$ with probability $|\alpha_c|^2$.

Sufficient condition for checking whether a QTM is *well-formed* – that is it satisfies requirements of quantum mechanics, like unitary evolution – were introduced by Bernstein, Vazirani and Hirvensalo.

# 3. CLASSICAL CELLULAR AUTOMATA

## 3.1 WHY CELLULAR AUTOMATA?

In the late 1940s, Von Neumann set to answer the question of whether a machine can possibly fabricate machines as complicated as themselves. In an attempt to simplify the problem, he considered that machines or automatons are made up of a small number of standardized parts.

In his model, a complex reservoir full of floating machine parts is used. Von Neumann was able proof, using mathematical logic, that a system made of an automaton and a blueprint for building the automaton can self-replicate by: First making a copy of the blueprint and then use the blueprint's instructions for making a copy of the automaton. It is interesting to observe that the blueprint is analogous to DNA for self-reproduction.

Von Neumann needed a simpler model to formulate a more convincing and constructive proof. S. Ulam, his colleague at Los Alamos, suggested thinking in terms of an idealized space of cells that hold finite states, where each state represent different machine part.

The idea was clearly presented in a paper by Ulam:

"*Given is an infinite lattice or graph of points, each with a finite number of connections to certain of its "neighbors". Each point is capable of a finite number of "states". The states of neighbors at time tn induce, in a specified manner, the state of the point at time $t_{n+1}$*"

Cellular Automata where used by von Neumann described a self reproducing machine that used 29 states, this work was completed in before 1952 but it was not published during his lifetime.

In 1970, Conway presented the Game of Life, which is a two-dimensional cellular automaton. In Life cells are either alive or dead, the rules are: a dead cell with exactly three live neighbors becomes alive. A live cell with two or three live neighbors stays so (survival). In all other cases, a cell dies or remains dead (overcrowding or loneliness). The game's simple rules generate beautiful patterns that to a certain extent seem 'alive'. Conway later proved that the Game of Life is computationally universal.

In the late 70s, Fredkin proposed that the world we live in is a huge cellular automaton. Fredkin's thought that all physical quantities can be seen as packets of information that reside in a cellular automaton.

S. Wolfram entered the field of cellular automata in the early 1980s. After a series of discoveries about one dimensional cellular automata, Wolfram decided to retire from publishing and indulge in a private investigation of cellular automata. In 2002, and after 15 years, Wolfram publishes his work in a book entitled A New Kind of Science (NKS). The book has been very controversial.

It remains to point the interesting fact that many of the leading researchers in Quantum Computing where also leading researchers in Cellular Automata, notable Charles Bennett, Tommaso Toffoli, and Norman Margolus.

Applications of Cellular automata range from parallel computation, artificial life, image processing and image generation, modeling biological systems, simulations of chemistry, simulation of physics, turbulence, algorithm and hardware design, graphics, and art.

## 3. 2 DEFINITION AND CLASSIFICATIONS

A Cellular Automaton (CA) is a regular discrete infinite network of identical finite automaton called cells, the state of which changes on discrete time steps. The evolution of each cell is deterministic and depends upon a local finite number of so-called neighboring cells.

More formally, a Cellular Automata is a quadruple $\langle d, Q, N, \delta \rangle$, where $d \in Z^{+*}$ is its dimension, Q is the finite set of all possible states, $N \subset Z^d$ is a finite neighborhood, $\delta: Q^{|N|} \rightarrow Q$ is a local transition function.

The *cells* of the automaton are organized in a line and are indexed by the elements of Z. The neighborhood of a cell is the set of all cells in the network which will locally determine its evolution. Von Neumann's and Moore's neighborhoods are two examples of commonly studied neighborhoods in d = 2.

In general the neighborhood of a cell do not have to be 'near', however it has been shown that every d-dimensional neighborhood can be simulated by a d-dimensional nearest neighborhood.

The radius of a neighborhood is the distance between the cell and its furthest neighbor. A related problem would be to optimize the radius versus the number of states. It is interesting to note that for a realistic model of computation we do not consider CAs whose ratio between the radius and the number of states is exponential since in such a setting NP-complete problems can be solved in polynomial time.

The following is a simple example of 1 dimensional, 2 state with radius=1, rule:

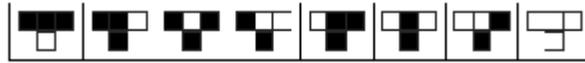

Bellow is a sketch of that simulates the evolution of the rule starting with an initial condition of {…00100…}:

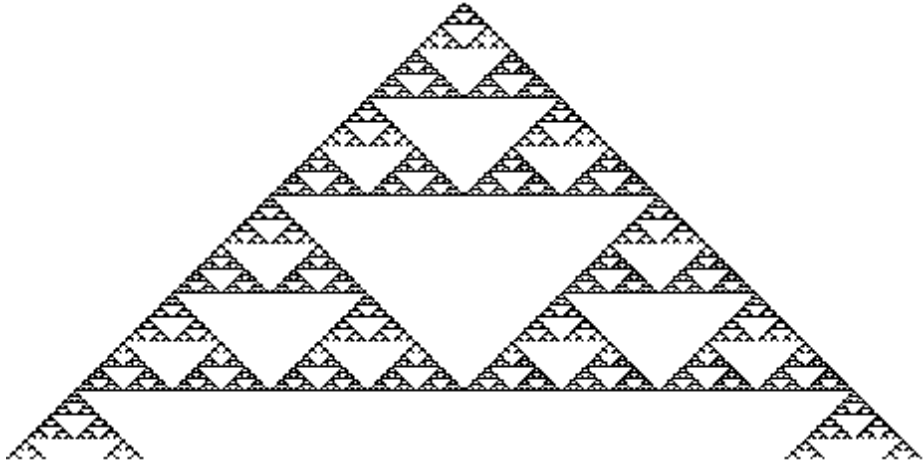

Wolfram devised an efficient representation of the local transition function $\delta$ (sometimes called rule). We describe the representation for the one-dimensional case:

There is $2^3=8$ different combination of states that determine the next state of a given cell, for every combination the rule can produce either a 0 or 1, so we have $2^8=256$ possible rules. Thus one can specify a rule by associating it with an 8-bit number determined by setting the ith bit to 0 if the corresponding combination outputs a 0 and 1 otherwise. For example the rule shown above is rule number 01111110=126.

In the early 1980s wolfram discovered that simple cellular automata rules can yield complex behavior, and he classifies cellular automata into four main categories:

Class 1 is characterized by simple behavior and that "information about initial conditions is always rapidly forgotten". Rule 254 is an example of this class; the following shows its evolutions starting with 6 different random initial conditions.

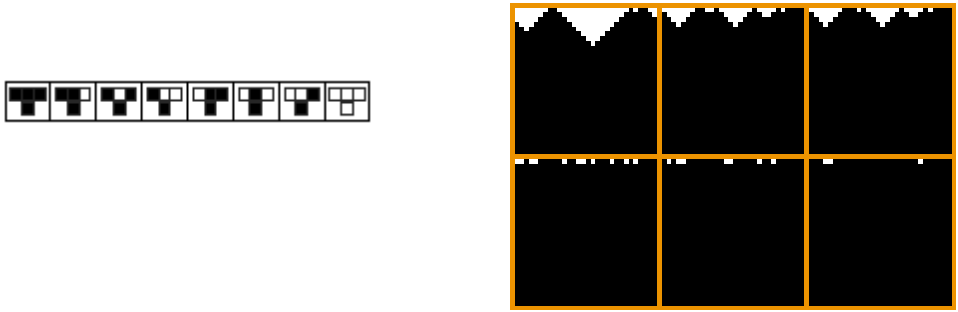

Class 2 is also simple but it has many fixed or periodic orbits, and from most initial states it will quickly converge to one of those orbits. In a Class 2 "some information in the initial state is retained in the final configuration…but this information always remains completely localized". Rule 170 is a common example:

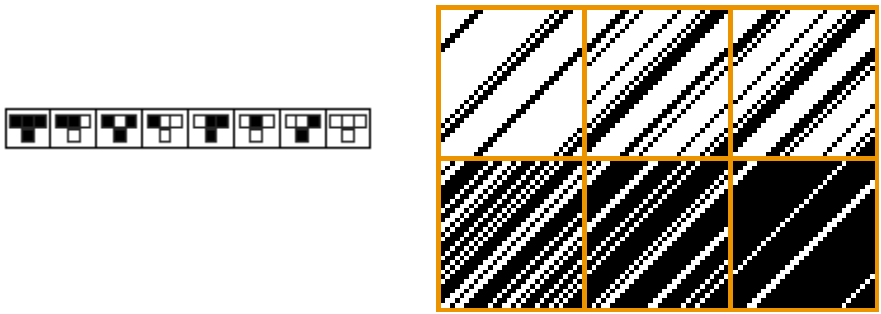

Class 3 behavior is characterized by apparent randomness it "shows long-range communication of information so that any change made anywhere in the system will almost always eventually be communicated even to the most distant parts of the system". It is somewhat surprising that there are simple cellular automata that belong to this class, Rule 30 is an example:

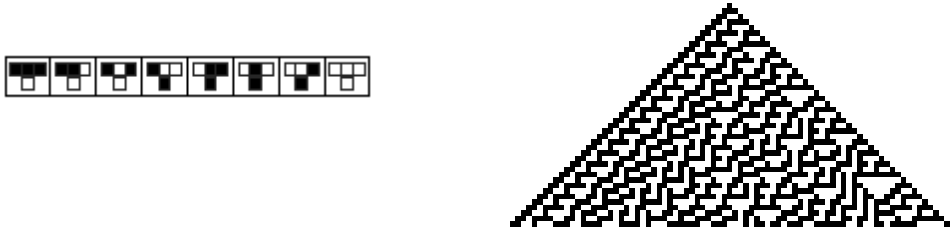

Class 4 is in-between Classes 2 and 3, it is characterized by regions with apparently random mixing and regions with localized structures that stay stationary or move linearly. A collision between two moving structures is an interaction that can be viewed as a two-bit a gate.

The Game of Life belongs to this class. Wolfram discovered that Rule 110 also belongs to this class and proved that it is computationally universal. Wolfram also conjunctures that all Class 4 cellular automata simulate the Turing Machine. In fact, Wolfram claims the principle of computational equivalence which states that "almost all processes that are not obviously simple can be viewed as computations of equivalent sophistication." The following shows Rule 110:

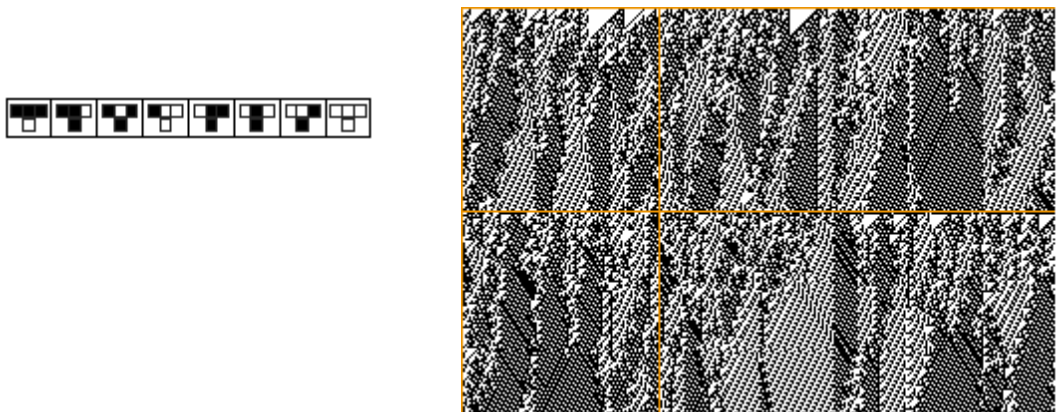

## 3. 3 PROPERTIES OF CELLULAR AUTOMATA

Back to general cellular automata $\langle d, Q, N, \delta \rangle$, one can formally describe the state of cellular automata at a given instant through a configuration (also called global state) which is mapping $c_t : Z_d \mapsto Q$, where $c_t \in Q^{z^d}$. The evolution of the cellular automaton is then a sequence ($c_t, _{t \geq 0}$) of configurations. Let us define the global transition function $c_i = G_\delta(c_{i-1})$ :

$G_\delta : Q^{z^d} \to Q^{z^d}$, and for every cell $\vec{Z}$ we have:

$G_\delta(c)(z) = \delta(c_t(z_1,...,z_d), c_t(z_1 + n_{11},...,z_d + n_{da}),...,c_t(z_1 + n_{1n},...,z_d + n_{dn}))$, where $((n_{11},...n_{da}),....,(n_{1n},...,n_{dn})) = (\vec{n_1},...,\vec{n_n})$ represent the neighborhood.

For example the von Neumann neighborhood in two has the following global transition function:

$G_\delta(c)(i, j) = \delta(c_t(i-1, j), c_t(i, j-1), c_t(i, j), c_t(i, j+1), c_t(i+1, j))$

A problem with the practical realization of general cellular automata is that an infinite array of cells is needed. There are two types of configurations that avoid this problem:

- Finite Configurations: These are configurations c, such that for a given quiescent state $q_e$, the support, defined as $Supp(c) = \{\vec{z} / c(\vec{z}) \neq q_e\}$, is finite.

- Periodic Configurations: These are configurations c, such that there exist a $\vec{p} \in Z^d$ such that $c(\vec{z}+\vec{p}) = c(\vec{z}), \forall \vec{z} \in Z^d$. These are suitable to describe automata on rings.

A cellular automaton is called:
- *Trivial* if each cell has only itself as a neighbor.
- *Simple* if its neighborhood is an interval of integers.
- *Symmetric* if the cell is the central element of its neighborhood.

An important subclass of cellular automata is reversible cellular automata, which was initially studied because cellular automata is used for to modeling and many phenomena are reversible.

A cellular automata is $A$ reversible if there is another cellular automata $A$' such that $G_\delta(c_1) = c_2$ if and only if $G_{\delta'}(c_2) = c_1$. Note that $A$' has a neighborhood that is in general different from $A$.

Hedlund and Richardson proved that a cellular automaton is reversible if and only if it is bijective.

Non-trivial reversible cellular automata are rare. In fact, for $d \geq 2$, it is undecidable whether a d-dimensional cellular automaton is reversible (Kari 1990). However, Toffoli (1977) proved that any k-dimensional CA can be simulated in real time by a (k+1)-dimensional reversible CA. Moreover, Dubacq (1985) proved that there is a universal cellular automaton that is reversible.

# 4. QUANTUM CELLULAR AUTOMATA

## 4.1 WHY QUANTUM CELLULAR AUTOMATA:

The very first attempts to create quantum versions of the classical models of computation have been oriented towards cellular automata. From an experimental point of view, QCA have a significant advantage over quantum circuits because cells are not required to be able to distinguish one neighbor from another. In cellular automata uniform rules are applied in parallel across a lattice, therefore it is not needed to address each qubit separately. This helps to eliminate error resulting from cross talk on neighboring qubits due to imperfectly aligned control fields. In addition, many fabrication techniques naturally produce equally spaced units suitable for cellular automata computation. Physical systems proposed for QCA include quantum dot arrays and endohedral fullerenes.

The first example of a QCA was Feynman's "quantum checkerboard" model of spinors in 2d spacetime, dating back at least to 1965.

The history of attempts to introduce the concept of quantum cellular automata goes back to Grossing and Zeilinger (1988) and a series of subsequent papers by the authors, however there model has little in common with the models currently in use.

In 1990, Norm Marglous wrote "Parallel Quantum Computation", later Biafore considered the problem of synchronization. The first successful model of one dimensional quantum cellular automata was due to Watrous (1995).

To some, studying QCA is theoretically motivated by the view that information is the fundamental element of nature and computing is the fundamental process. The universe is then perceived as giant quantum cellular automaton.

## 4. 2 DEFINITION AND RESTRICTIONS:

Watrous's model deals with quiescent QCA that have finite configurations:

A 1d-QCA is a quadruple $\langle Q, \lambda, N, \delta \rangle$, Q is the finite set of all possible states including the quiescent state $\lambda$, $N = \{n_1, ...., n_r\}$ is a finite neighborhood, and $\delta: Q^{r+1} \to C_{[0,1]}$ is a local transition function.

Where $C_{[0,1]}$ is an interval in the field of complex number of norm in [0,1]. To see how the transition function works consider an example:

If we are working with a symmetric quantum cellular automata with a neighborhood of size 3 then if three consecutive cells are in the states $q_1$, $q_2$, and $q_3$ at time t, then at time t+1 the middle cell will contain every $q \in Q$ with amplitude $\delta(q_1, q_2, q_3, q)$.

Again the concept of a configuration $c_t : Z \mapsto Q$ is used to define how a 1d-QCA evolves. Since we are working with finite configurations, a valid configuration is such that $c(i) \neq \lambda$ for only finitely many $i$.

If we let C denote the set of all such configurations, then computation is done in the inner product space $H = l_2(C)$ with bases $\{|c\rangle | c \in C\}$. The evolution of 1d-QCA can then be though of as every configuration $c_1$ is transferring to multiple configurations with corresponding weights, the amplitude of the transition to configuration $c_2$ is the product of amplitudes with which each cell of $c_1$ transforms to the corresponding cell in $c_2$:

$$\alpha(c_1, c_2) = \prod_{i \in Z} \delta(c_1(i+n_1), c_1(i+n_2), ..., c_1(i+n_r), c_2(i))$$

So in general a state in $H$ has the form $|\Phi\rangle = \sum_{c \in C} \alpha_c |c\rangle$, where $\sum_{c \in C} |\alpha_c|^2 = 1$.

The evolution operator is then defined to be:

$$|\Psi\rangle = E|\Phi\rangle = \sum_{c \in C} \beta_c |c\rangle \text{ where } \beta_c = \sum_{c' \in C} \alpha_{c'} \alpha(c', c).$$

δ has to satisfy the following conditions:

- Local probability condition: for any $(q_1, ..., q_r) \in Q^r$ then

$$\sum_{q \in Q} |\delta(q_1, ..., q_r, q)|^2 = 1$$

- Stability of the quiescent state condition: if $q \in Q$ then $\delta(\lambda, ..., \lambda, q) = 1$ if $q = \lambda$ and 0 otherwise.

- Unitarily if evolution condition: the mapping $E$ has to be unitary.

Verifying that a QCA is unitary is generally not easy. A QCA is said to be *well-formed* if the evolution operator properly transforms probabilities by preserving there sum-squared to 1. An alternative definition is to say that $E$ preserves the $l_2$- norm (equivalently $E$ has to be injective). Durr, LeThanh and Santha devised an algorithm to check this property in $O(n^2)$ time.

Well-formedness is equivalent to the orthonormality of the column vectors of the time evolution operator; unitarity also requires orthonormality from the row vectors ($E$ has to be Bijective).

Durr and Santha published an algorithm that checks in $O(n^{\frac{3r+1}{r+1}})$ time if a 1d-QCA is unitary, where r is the size of the neighborhood.

A trivial QCA is unitary if and only if $\forall q_1, q_2 \in Q$:

$$\sum_{q \in Q} \delta(q_1, q)\delta(q_2, q) = 1 \text{ if } q_1 = q_2, \text{ 0 otherwise.}$$

Generalizing the definition to higher dimensional QCA is not straight forward since for $d \geq 2$, it is undecidable whether a d-dimensional QCA is valid. This result directly follows from Kari's theorem for reversible CA.

In the literature, quiescent QCA is the most common, periodic QCA where considered by Van Dam in his master's thesis. Van Dam showed that for periodic QCA there exists a universal quantum cellular automaton (in the sense that it can simulate any other cellular automaton in polynomial steps). This result does not apply, however, to the more general quiescent QCA. In fact Durr has proved that periodic QCA are a subset of quiescent QCAs.

4. 3 SUBCLASSES OF QCA

Partitioned Cellular Automata (PCA) is a special case of CA introduced by Morita and Harao in 1989. Watrous introduced the quantum version since it is easy to determine whether a PQCA is unitary or not.

A one dimensional Partitioned Quantum Cellular Automata is a 1d-QCA in which each cell is into three sub-cells: left, middle, and right. The next state of a cell depends on the state of: the right sub-cell of the left neighbor, the middle sub-cell of itself, and the left sub-cell of the right neighbor.

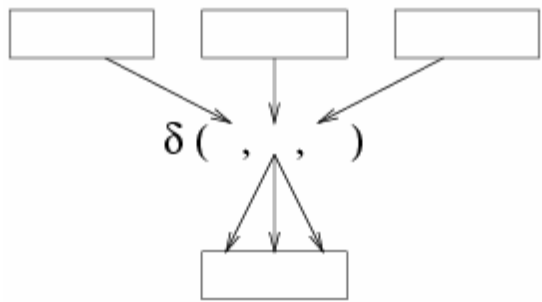

In general a 1d-PQCA is a 1d-QCA quadruple $\langle Q, \lambda, N, \delta \rangle$ with the following properties:

- The set of states is Cartesian product $Q = Q_1 \times ... \times Q_r$ of r sets.
- The local transition function $\delta$ is a composition of two functions:
  - $\delta_c : Q^r \to Q$:

    $\delta_c((q_{1,1},...,q_{1,r}),(q_{2,1},...,q_{2,r}),....,(q_{r,1},...,q_{r,r})) = (q_{1,1}, q_{2,2},...,q_{r,r})$

    is the permutation on the sub-cells in the neighborhood of a cell, the effect of $\delta_c$ for r = 3 is shown bellow:

    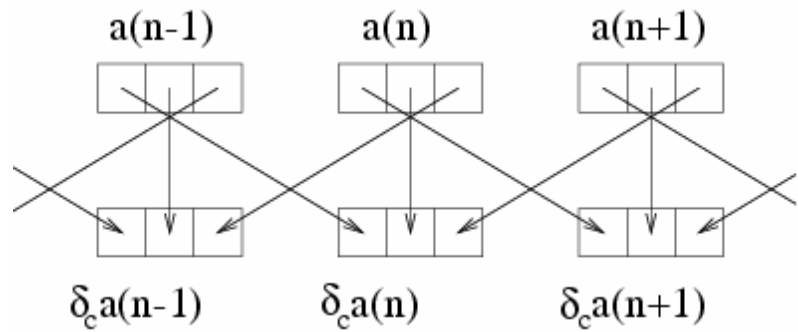

  - $\delta_q : Q \to C^Q$ is a quantum mapping, $\delta_c$ can be considered as the transition function a trivial QCA.

PQCA can be seen as concatenation of a classical CA and a QCA. The role of classical CA is permuting the sub-cells and thus allowing the cells to 'communicate' between one another. The trivial QCA, in turn, is what makes PQCA quantum and allows it to perform quantum computation.

One can also view the classical CA permutation performed by $\delta_c$ to represent a 1d-QCA with the evolution operator satisfies:

$U(q_1,q_2) = 1$ if $G_{\delta_c}(q_1) = q_2$, 0 otherwise.

Since $U$ is unitary, the evolution of 1d-PQCA is unitary if and only if the evolution of the corresponding trivial QCA described by $\delta_c$ is unitary.

The following is example is taken from Gruska's excellent book "Quantum Computing":

A technical report by Watrous in 1997 presents an example of a simple PQCA $A = \langle Q, \lambda, U \rangle$ that simulates, in a sense, the EPR phenomenon. Let,

$Q = Q_l \times Q_m \times Q_r$, where $Q_l = Q_r = \{0,+,-\}, Q_m = \{0\}, \lambda = (0,0,0)$.

The matrix U of degree $|Q|$ is defined by

$$U(q',q) = \begin{cases} 1, & \text{if } q = q' \notin \{(-,0,+),(+,0,-)\}; \\ 0, & \text{if } q \neq q' \text{ and } \{q,q'\} \neq S, \end{cases}$$

and for $q,q' \in S$,

$$U(q',q) = \begin{cases} -\dfrac{1}{\sqrt{2}}, & \text{if } q = q' = (-,0,+); \\ \dfrac{1}{\sqrt{2}}, & \text{otherwise.} \end{cases}$$

If A evolves from the initial basis state $|c\rangle$, where c is defined by

$$c(n) = \begin{cases} (0,0,-), & \text{if } n = -1; \\ (+,0,0), & \text{if } n = 1; \\ \lambda, & \text{otherwise,} \end{cases}$$

then, after one step, A is in the superposition

$$\frac{1}{\sqrt{2}}(|c_0\rangle + |d_0\rangle)$$

where

$$c_0(n) = \begin{cases} (+,0,-), & \text{if } n = 0; \\ (0,0,0), & \text{otherwise,} \end{cases} \qquad d_0(n) = \begin{cases} (-,0,+), & \text{if } n = 0; \\ (0,0,0), & \text{otherwise,} \end{cases}$$

By induction one can show that after $t > 1$ steps A is in the superposition

$$\frac{1}{\sqrt{2}}(|c_{t-1}\rangle + |d_{t-1}\rangle),$$

where

$$c_t(n) = \begin{cases} (+,0,0), & \text{if } n = -t; \\ (0,0,-), & \text{if } n = t; \\ (0,0,0), & \text{otherwise,} \end{cases} \qquad d_t(n) = \begin{cases} (-,0,0), & \text{if } n = -t; \\ (0,0,+), & \text{if } n = t; \\ (0,0,0), & \text{otherwise,} \end{cases}$$

If we interpret the states $(0,0,-)$ and $(-,0,0)$ as negative particles and $(0,0,+)$, $(+,0,0)$ as positive particles, then the configuration $|c_t\rangle$ models the situation in which the positive particles moves to the left and the negative ones to the right; the configuration $|d_t\rangle$ models the situation when particles are reversed.

Another subclass of QCA is Block-partitioned QCA (BQCA), the classical version was introduced by Margolus and Toffoli in 1980s.

A particular BQCA is a QCA whose cells are divided into blocks of two. The rule maps block states into block states. Notice that this is different from other cellular automata in which the mapping is into individual cells not groups of cells.

In order to allow 'communication' between different parts of the cellular automaton we let the blocks shift to the right at every time step, alternatively we can think of them shifting to the left and then back to right and so forth.

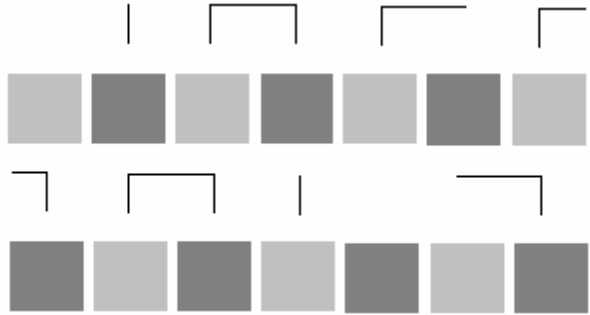

The rules are applied uniformly to every block; It is sometimes allowed to change rule that is applied at each time-step. The rules can now be viewed as two qubits gates applied to the cells at positions $2i$ and $2i+1$ on even time steps and to positions $2i+1$ and $2i+2$ on odd time steps. BQCA are very popular, it was recently proposed to use in order explore the raw computational properties of a physical system.

A Quantum Gate Cellular Automata evolves by a repeated sequence of two steps: one step acts to permute the sub-cells, and the second step applies parallel gates over the neighborhood.

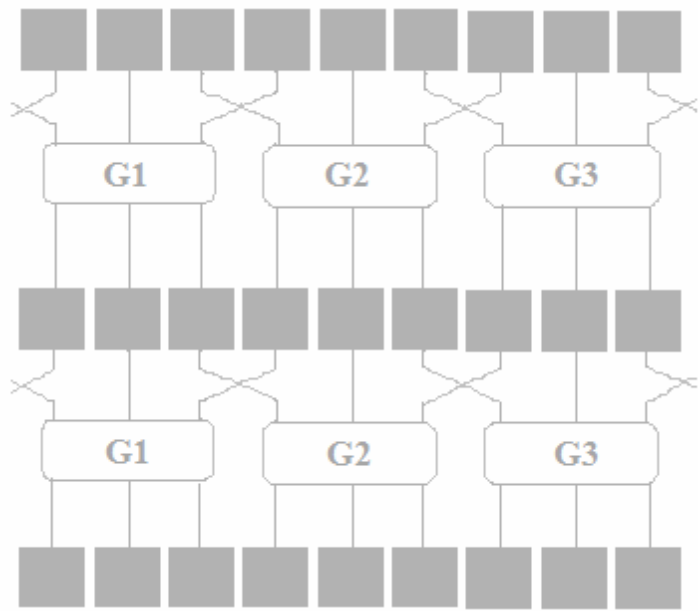

It is useful to compare QGCA and BQCA:
- In BQCA the same gate is applied to all the pairs in the entire cellular automaton at a given time, the gate may be different in different times.
- In QGCA the different gates are applied to different pairs in the cellular automaton at a given time, the gates do not change with time.

## 4.4 UNIVERSALITY

The mutual efficient simulation of QTMs and QCAs is not fully understood, however preliminary results suggest that QCAs are much more powerful than QTMs.

Watrous (1996) has shown that:

- PQCA can simulate any QTM with constant slowdown.
- QTMs can simulate any PQCA with a linear slowdown.

A QTM *M* accepts an input $x \in \Sigma^*$ if the tape square *k* contains $a \in \Sigma_a$ after processing is completed, where *k* is a distinguished tape square and $\Sigma_a$ is a set of acceptance states. Similarly a 1d-QCA *A* accepts if the cell number *k* reaches a state $a \in Q_a$ after some evolution starting at a configuration $x \in Q^*$, where *k* is a distinguished cell and $Q_a$ is a set of acceptance states.

To illustrate Watrous's proof of the first statement we introduce Unidirectional Quantum Turing Machines (uQTMs):

They are QTMs in which a state specifies a direction that restrict reaching it except if the head is moving in this direction. Formally, $d_1 = d_2$ whenever $\delta(q_1, \sigma_1, \sigma`_1, q, d_1) \neq 0$ and $\delta(q_2, \sigma_2, \sigma`_2, q, d_2) \neq 0$. Bernstein and Vazirani proved that any QTM can be simulated with uQTM with a constant slow down.

Therefore if construct a PQCA capable of simulating a uQTM then we would also have a PQCA simulating a general QTM. A PQCA $A$ simulating a uQTM $M$ can now be defined as follows:

The state space of a cell is of the form $Q = Q_l \times Q_m \times Q_r$ such that $Q_l = K_l \cup \{\#\}$, $Q_m = \Sigma$, and $Q_r = K_r \cup \{\#\}$, where $\# \notin K$, $\lambda = (\#, b, \#)$ and $b$ is the blank symbol. The transition matrix $U$ is defined as follows:

1. For each $(s_1, \tau_1)$, $(s_2, \tau_2) \in K_l \times \Sigma$ let $U((s_2, \tau_2, \#), (s_1, \tau_1, \#)) = (s_1, \tau_1, \tau_2, s_2, \leftarrow)$.

2. For each $(s_1, \tau_1) \in K_l \times \Sigma$, $(s_2, \tau_2) \in K_r \times \Sigma$ let

   $U((\#, \tau_2, s_2), (s_1, \tau_1, \#)) = (s_1, \tau_1, \tau_2, s_2, \rightarrow)$,

   $U((s_2, \tau_2, \#), (\#, \tau_1, s_1)) = (s_2, \tau_2, \tau_1, s_1, \leftarrow)$

3. For each $(s_1, \tau_1)$, $(s_2, \tau_2) \in K_r \times \Sigma$ let $U((\#, \tau_2, s_2), (\#, \tau_1, s_1)) = (s_1, \tau_1, \tau_2, s_2, \rightarrow)$.

4. For any $q_1, q_2 \in Q$ for which $U(q_1, q_2)$ has not been defined in (1) to (3), let $U(q_1, q_2) = 1$ if $q_1 = q_2$, 0 otherwise.

Let $k_A = k$ and $Q_a = \{(q_l, q_m, q_r) \in Q \mid q_m \in \Sigma_a\}$, it is straightforward to show that that if $M$ is a valid (well-formed) QTM then $A$ is unitary.

Finally we need to define the mapping $T: C_M \rightarrow C_A$:

$T(xqy)(n) = (\#, xy_n, \#)$ if $q \in K_l, n \neq |x| + 2$

$\phantom{T(xqy)(n) =} (q, xy_n, \#)$ if $q \in K_l, n = |x| + 2$

$\phantom{T(xqy)(n) =} (\#, xy_n, \#)$ if $q \in K_r, n \neq |x|$

$\phantom{T(xqy)(n) =} (\#, xy_n, q)$ if $q \in K_r, n = |x|$

Where $xy_n$ denotes the $n$th symbol of the string $xy$, or $\lambda$ if out of range.

It is easy to check that the probability that $M$ accepts the initial configuration $q_o w$ after n steps is equal to the probability that $A$ accepts $T(q_o w)$ after n steps.

This completes Watrous's proof that PQCA can simulate any QTM with constant slowdown.

It is not known whether a single BQCA rule could simulate the QTM; however it can be shown that a sequence of rules can simulate a quantum computer with a linear slowdown. This follows from Benjamin's results that an open 1D lattice composed of an alternating array of two species ABAB… of qubits can be used for quantum computation.

4.5 A SIMPLE UNIVERSAL ARCHITECTURE

The architecture used by Benjamin is a one dimensional array of two state cells labeled in pattern *ABAB*… One cell at one end of the array is "special" in that it might be independently controlled and associated with a measuring device.

For the remaining n qubits, the Hamiltonian of the is given by $H_{total} = \sum_{i=1}^{n} H_i^s + \sum_{i=1}^{n-1} H_{i,i+1}^{int}$ where the $H_i^s$ represents the energy of each qubit in isolation, and $H_{i,i+1}^{int}$ represents the interaction between neighboring qubits. The Hamiltonians are restricted in that for all $i$, $H_{2i}^s = H^A, H_{2i+1}^s = H^B, H_{2i,2i+1}^{int} = H^{AB}$ and $H_{2i-1,2i}^{int} = H^{AB}$. For universal computation the Hamiltonian must satisfy:

(1) The terms $H^{AB}$ can be "switched off" such that the system then decouples into a set of *A*-*B* pairs.

(2) By manipulation of the remaining terms $H^A$, $H^B$, and $H^{BA}$, it must be possible to implement any two-qubit operation on the *A*-*B* pairs

The same conditions apply for $H^{BA}$. Given that any two-qubit operation is allowed then one sets up 1s and 0s then uses SWAPs and controlled-$U$ transformations to apply $U$ to individual qubits. It can also be shown that scheme implements individual two-qubit operations.

This architecture can be implemented using quantum dots. Benjamin speculated on several systems in the solid state where qubits are represented by the spins of individual electrons, whose wave functions (light gray) are trapped in confinement potentials (dark gray):

(a) Confining potentials are shaped so that there is only significant wave function overlap when there is an applied field (left). Moreover, the sign of the field determines whether the qubits couple in the pattern $AB, AB \ldots$, or the pattern $\ldots BA, BA \ldots$. An alternative (right) based on a sharp confinement potential alternating with an elongated one.

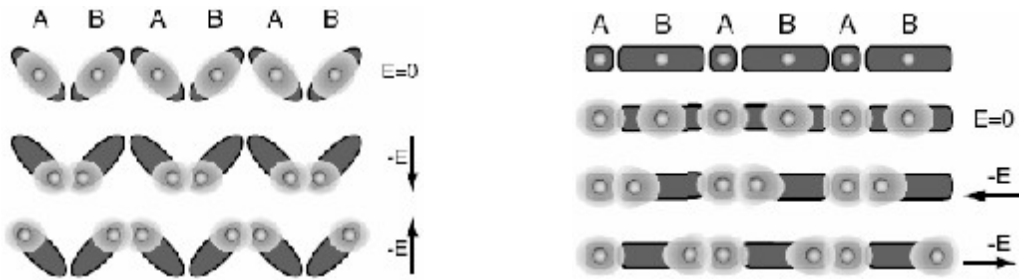

(c) One can also use the classical method of electrodes to control the qubit-qubit interactions, but because the interaction is switched collectively, two electrodes for the entire $n$ qubit array.

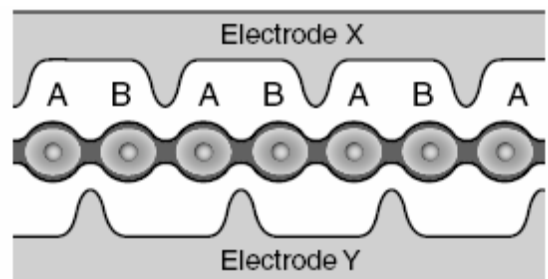

## 5. ADVANCED ISSUES

Recent results in Quantum Cellular Automata:

Benjamin and Johnson presented the architecture constitutes a novel paradigm whereby the algorithm is embedded in spatial, as opposed to temporal, structure. The architecture was used to search in time of O($N^{1/3}$).

Gavin K. Brennen and Jamie E. Williams generalized BQCA to open quantum systems, which evolve according to non-unitary rules. The BQCA rule is then represented by an appropriate set of Krauss operators acting on the system density matrix. This more generalized treatment can be thought of as including the effect of correlated noise in the quantum evolution.

Michael McGuigan generalized the definition of quantum cellular automata to include bosonic, fermionic, and supersymmetric and spin quantum system, by applying methods of lattice field theories to the quantization of the bosonoic and fermionic models using the path integral formulation.

END OF PAPER            END OF PAPER            END OF PAPER